\documentclass[a4paper,11pt]{article}
\pdfoutput=1
\usepackage[utf8x]{inputenc}
\usepackage{jcappub}
\usepackage{cleveref}
\usepackage{amstext,amssymb}
\usepackage{amsmath}
\usepackage{graphicx}
\usepackage{xspace}
\usepackage{color}
\usepackage{units}
\usepackage{slashed} % feynman slash notation via \slashed{p}
\usepackage{multirow}
\usepackage{subfig}
\newcommand{\be}{\begin{equation}}
\newcommand{\ee}{\end{equation}}
\newcommand{\bea}{\begin{eqnarray}}
\newcommand{\eea}{\end{eqnarray}}
\newcommand{\nn}{\nonumber}

\def\s1{\hat s}

\usepackage{slashed}
\global\long\def\d{\partial}

\begin{document}
%\title{\Large Majorana Dark Matter in a new $B-L$ model}
\title{ Majorana Dark Matter in a new $B-L$ model}
\author[a]{Shivaramakrishna Singirala}
\author[a]{Rukmani Mohanta}
\author[b]{Sudhanwa Patra}
\author[c]{Soumya Rao}

\affiliation[a]{School of Physics,  University of Hyderabad, Hyderabad - 500046,  India}
\affiliation[b]{Indian Institute of Technology Bhilai, GEC Campus, Sejbahar, Raipur-492015, Chhattisgarh, India}
\affiliation[c]{National Centre for Nuclear Research, Ho{\.z}a 69, 00-681 Warsaw, Poland}

\emailAdd{krishnas542@gmail.com}
\emailAdd{rmsp@uohyd.ernet.in}
%\affiliation{School of Physics,  University of Hyderabad, Hyderabad - 500046,  India}
\emailAdd{sudhanwa@iitbhilai.ac.in}
\emailAdd{somu.jrf@gmail.com}

% \author[c]{S. Econd,}

\abstract{
We present a comprehensive study of Majorana dark matter in a $U(1)_{B-L}$ gauge extension
of the standard model, where three exotic fermions with  $B-L$ charges as $-4, -4, +5$
are added to make the model free from the triangle gauge anomalies. The enriched scalar
sector and the new heavy gauge boson $Z^{\prime},$ associated with the $U(1)_{B-L}$
symmetry make the model advantageous to be explored in dual portal scenarios for the
search of dark matter signal. 
Diagonalizing the exotic fermion mass matrix, we obtain the Majorana mass eigenstates, of which the
lightest one plays the role of dark matter. Analyzing the effect of two mediators
separately, the scalar portal channels give a viable parameter space consistent with relic
density from PLANCK data and the direct detection limits from various experiments such as
LUX, XENON1T and PandaX. While the $Z^{\prime}$ mediated channels are
constrained from relic abundance and LHC searches for $Z^{\prime}$ in the dilepton
channel. A massless physical Goldstone boson plays a key role in the scalar portal relic density. Finally, we briefly discuss the neutrino mass generation at one-loop level.
 
}

%%%%%%%%%%%%%%%%%%%%
% \maketitle
%\pacs{95.35.+d, 11.15.Ex}
\maketitle
\flushbottom
\section{Introduction} 
Standard model (SM) of particle physics is the most   successful theory that can  explain well
almost all the observed data  below the electroweak scale. Still there are many open issues for
which SM does not provide any satisfactory answer. Of these open questions, the one that stands out is
the nature of  Dark matter (DM).  Various observational evidences firmly point towards the existence
of dark matter which constitutes about $26.8\%$ energy budget of the Universe \citep{Ade:2015xua},
still very little is known about its true nature. The DM has been the most sought after candidate
for experimental particle physicists and the  hunt for it started from the day of its
existence, which was proposed way back in 1937 \cite{Zwicky:1937zza,Rubin:1970zza}. Its nature to interact
`weakly' as  confirmed indirectly from the Bullet cluster \cite{Clowe:2003tk}, provides strong
motivation to prefer weakly interacting massive particles (WIMPs) as the potential DM candidates,
which are not too far from the electroweak scale, thus, providing an excellent testing ground at the
current or near future direct or indirect dark matter detection experiments. 

To explain the key ingredient that connects cosmology with the particle physics, plenty of
frameworks have been  proposed imposing the condition that the DM is stable at the
cosmological time scale. Apart from this, as the spin of DM is unknown, all possible kinds
of DM candidates, i.e., scalar, fermion, vector have been explored.  As SM is a well
tested gauge theory, we intend to study the gauge extensions of it where the difference of
Baryon and Lepton number $(B-L)$  is promoted to the local gauge symmetry
\cite{Jenkins:1987ue,Buchmuller:1991ce,Basso:2008iv,Emam:2007dy}.  One of the interesting
aspects is that in its standard form, the presence of right-handed neutrinos and the
type-I seesaw mechanism for neutrino mass generation is natural.  In particular, $B-L$
gauge extension of SM has been studied so as to incorporate the beyond Standard Model
(BSM) physics (see some earlier works in this
motivation~\cite{Khalil:2006yi,Iso:2009ss,Kanemura:2014rpa,
Lindner:2011it,Okada:2016gsh,Okada:2016tci,Bhattacharya:2016qsg,
Biswas:2016ewm,Wang:2015saa,Basak:2013cga,Bandyopadhyay:2017bgh,
DeRomeri:2017oxa,Okada:2010wd,Kanemura:2011vm,Seto:2016pks}).  In this article, we explore the prospects
for Majorana DM in the context of $B-L$ gauge extensions of SM.

The model considered here consists of a particular $B-L$ charge assignment for the extra
fields added to SM, such that there is an automatic cancellation of anomalies as well as
the existence of a stable Majorana DM
candidate~\cite{Ma:2014qra,Ma:2015raa,Ma:2015mjd,Patra:2016ofq}.  The model incorporates a
scalar sector with two additional heavy scalars alongside the SM Higgs.  In particular,
one of the scalars carries a $B-L$ charge of $+8$ which gives rise to a scalar portal
interaction with the Majorana DM candidate such that there can be observable signals.
This is complemented by the usual $Z^{\prime}$ mediated interactions, where the Majorana
DM couples to the $Z^{\prime}$.  Thus, we have a Majorana DM
that can interact with SM particles through two portals - one scalar and one vector.  We
shall study the phenomenology resulting from both these types of interactions using
constraints from direct and indirect detection of DM, as well as collider searches for
$Z^{\prime}$.  

%We also look at the possibility of explaining the galactic center
%$\gamma$-ray excess in the scalar mediated channel, particularly near the Higgs resonance.

The paper is organized as follows. In section II, we discuss a new variant of
$U(1)_{B-L}$ gauge extension of SM with three exotic fermions and an extended scalar
sector.  In section III, we give details of the masses and mixings in the fermion and scalar
sectors after spontaneous symmetry breaking.  Section IV discusses the dark matter phenomenology including
relic density and direct constraints.  Collider limits
on the current model are investigated in section V. Discussion regarding the generation of neutrino mass is presented in section VI and finally we conclude in section VII.

\section{New $B-L$ model with Majorana Dark Matter}
We consider an anomaly free $U(1)_{B-L}$ gauge extension of the SM where three exotic neutral
fermions with $B-L$ charges $-4, -4, +5$ are added to get rid of the non-trivial triangle gauge
anomalies. This minimal charge assignment was first proposed in \cite{Montero:2007cd} and later explored in \cite{Ma:2014qra,Ma:2015raa}. Another possibility is to add four exotic fermions charged $4/3,1/3,-2/3$ and$-2/3$ under new $U(1)$, which was first put forth in Ref.~\cite{Patra:2016ofq} and later studied in dark matter context in \cite{Nanda:2017bmi}.  However, we do not study the second possibility as it has already been studied in the dark matter context and due to the fact that our present choice requires the addition of only three fermions.
%In addition, two extra scalar singlets $\phi_1$ and $\phi_8$ are introduced to spontaneously break the $SU(2)_L \times U(1)_Y \times U(1)_{B-L}$ to $SU(2)_L \times U(1)_Y$ and further broken by SM Higgs doublet $H$ to $SU(2)_L \times U(1)_{em}$. 
In  addition, two scalar singlets $\phi_1$ and $\phi_8$ are introduced to generate the mass terms for the exotic neutral fermions after the spontaneous breaking of $B-L$ gauge symmetry.
Singlet dark matter in the
similar context has been explored recently in \cite{Singirala:2017see}. 
%================================================
\begin{table}[htb]
\begin{center}
%================================================
\begin{tabular}{|c|c|c|c|}
	\hline
			& Field	& $ SU(2)_L\times U(1)_Y$	& $U(1)_{B-L}$	\\
	\hline
	\hline
	Fermions	& $Q_L \equiv(u, d)^T_L$			& $(\textbf{2},~ 1/6)$	& $1/3$	\\
			& $u_R$							& $(\textbf{1},~ 2/3)$	& $1/3$	\\
			& $d_R$							& $(\textbf{1},~-1/3)$	& $1/3$	\\
			& $\ell_L \equiv(\nu,~e)^T_L$	& $(\textbf{2},~  -1/2)$	&  $-1$	\\
			& $e_R$							& $(\textbf{1},~  -1)$	&  $-1$	\\
			& $N_{1R}$						& $(\textbf{1},~   0)$	&  $-4$	\\
			& $N_{2R}$						& $(\textbf{1},~   0)$	&  $-4$	\\
			& $N_{3R}$						& $(\textbf{1},~   0)$	&  $5$	\\
	\hline
	Scalars	& $H$							& $(\textbf{2},~ 1/2)$	&   $0$	\\
			& $\phi_1$						& $(\textbf{1},~   0)$	&  $-1$	\\  
			& $\phi_8$						& $(\textbf{1},~   0)$	&  $8$	\\  
	\hline
	\hline
\end{tabular}
%================================================
\caption{Fields and their charges of the proposed $U(1)_{B-L}$ model.}
\label{tab:BL}
\end{center}
\end{table}
%================================================
Using the particle content listed in Table \,\ref{tab:BL}, one can write the following invariant Lagrangian
\begin{eqnarray}
\label{eq:TheModel}
	\mathcal{L}_\text{BL} 
	&&= -\frac{1}{3} \,g_\text{BL}\overline{Q}_L \,Z_\mu^\prime \gamma^\mu Q_L   
            -\frac{1}{3}\,g_\text{BL} \overline{u}_R \,Z_\mu^\prime \gamma^\mu u_R
            -\frac{1}{3} \,g_\text{BL}\overline{d}_R \,Z_\mu^\prime \gamma^\mu d_R  \nonumber \\
        &&~~~    + \,g_\text{BL}\overline{\ell}_L \,Z_\mu^\prime \gamma^\mu \ell_L
            + \,g_\text{BL} \overline{e}_R \,Z_\mu^\prime \gamma^\mu e_R      
      + i \, \overline{N}_{1R} \left( \slashed{\d} + 4i\,g_\text{BL} \,Z_\mu^\prime \gamma^\mu \right)\,N_{1R} \nonumber\\
      &&~~~	     + i \, \overline{N}_{2R} \left( \slashed{\d} + 4i\,g_\text{BL} \,Z_\mu^\prime \gamma^\mu \right)\,N_{2R}	
	      + i \, \overline{N}_{3R} \left( \slashed{\d} - 5i\,g_\text{BL} \,Z_\mu^\prime \gamma^\mu \right)\,N_{3R} 
	\nonumber \\
	&&~~~ - \frac{y_{\alpha \beta}}{2} \left( \sum_{\alpha, \beta=1,2}\overline{N^c_{\alpha R}}\,N_{\beta R} \,\phi_8 + h.c.\right) 
          - \frac{y_{\alpha 3}}{2} \left( \sum_{\alpha=1,2}\overline{N^c_{\alpha R}}\,N_{3 R} \,\phi_1 + h.c.\right) 
%	     - \frac{y_D}{2} \left( \bar{L}\,\tilde{H} \nu_R +h.c. \right) 
	\nonumber \\ 
	&&~~~ + |\left( \d_\mu + \,i\,g_\text{BL}\,Z'_\mu \right) \phi_1|^2
		 + |\left( \d_\mu -8 \,i\,g_\text{BL}\,Z'_\mu \right) \phi_8|^2 
	\nonumber \\
	&&~~~ - \frac{1}{4} F_{Z^\prime}^{\mu\nu}F^{Z^\prime}_{\mu\nu} 
	     + \frac{\kappa}{4} F_{Z^\prime}^{\mu\nu}\,F_{\mu\nu}
	     - V \left( H, \phi_1,\phi_8\right)
	     + \mathcal{L}_\text{SM} 
	  \, ,
\end{eqnarray}
where $Z_\mu^{\prime}$ is  the new gauge boson associated with $B-L$ gauge symmetry. 
Also $F_{\mu\nu}^{Z^\prime} = \partial_\mu Z_\nu^\prime - \partial_\nu Z_\mu^\prime$ is the corresponding 
field strength tensor for $U(1)_{B-L}$. The term containing $\kappa$ is the kinetic mixing term 
between the two $U(1)$ gauge groups. However, electroweak measurements severely constrain the corresponding mixing angle to be $\leq 10^{-3}$ \cite{Erler:2009jh}. In the present work we neglect this small mixing.

The scalar potential of the
model is given by
 \begin{align}
V(H,\phi_1,\phi_8)= & \mu^2_H  H^\dagger H + \lambda_H (H^\dagger H)^2  + \mu^2_1 \phi^\dagger_1 \phi_1 + \lambda_1 (\phi^\dagger_1 \phi_1)^2 
      +\mu^2_8 \phi^\dagger_8 \phi_8  \nonumber \\
      +& \lambda_8 (\phi^\dagger_8 \phi_8)^2 + \lambda_{H1} (H^\dagger H) (\phi^\dagger_1 \phi_1)
      +\lambda_{H8} (H^\dagger H) (\phi^\dagger_8 \phi_8)\nonumber \\
     + &  \lambda_{18} (\phi^\dagger_1 \phi_1) (\phi^\dagger_8 \phi_8).
      \label{scalarpot}
\end{align} 
%The scalar potential (\ref{scalarpot}) respects two
%	accidental global $U(1)$ symmetries associated with $\phi_1$ and $\phi_8$. Once
%	they are spontaneously broken by the vevs of $\phi_1$ and $\phi_8$, two massless
%	Goldstone modes arise in the model.

The stability of the scalar potential of the model is guaranteed by the co-positive criteria given by
$\lambda_H \ge 0$, $\lambda_1 \ge 0$, $\lambda_8 \ge 0$, $\lambda_{H1} + \sqrt{\lambda_H
\lambda_1} \ge 0$, $\lambda_{H8} + \sqrt{\lambda_H \lambda_8} \ge 0$, $\lambda_{18} +
\sqrt{\lambda_1 \lambda_8} \ge 0,  \sqrt{\lambda_H \lambda_1 \lambda_8} + \lambda_{H1}\sqrt{\lambda_8} + \lambda_{H8}\sqrt{\lambda_{1}} + \lambda_{18} \sqrt{\lambda_H} \ge 0$.
Tree level perturbative unitarity constrain the scalar couplings as 
\begin{eqnarray}
&&\lambda_{H},\lambda_{1},\lambda_8 \le 4\pi/3, \nn\\ 
&&\lambda_{H1}, \lambda_{H8},
\lambda_{18} \le 4\pi.\label{unitarity}
\end{eqnarray}

\section{Spontaneous symmetry breaking, masses and mixing}
The spontaneous symmetry breaking of $SU(2)_L \times U(1)_{Y} \times U(1)_{B-L}$ down to SM gauge group 
$SU(2)_L \times U(1)_Y$ is implemented with the scalars $\phi_1$ and $\phi_8$. Then the spontaneous symmetry 
breaking of SM gauge group to low energy theory is achieved by assigning a non-zero VEV to SM Higgs doublet.  
%This is a consequence of
%the choice of exotic charges for the Majorana fermions in the model.
%The non-zero VEVs of these scalars provide masses for fermions and gauge bosons.  
	Similar kind of $B-L$ model with additional scalars with $\phi_1$ and $\phi_2$ has
	been discussed in Ref~\cite{Patra:2016ofq},  which avoids the presence of
	any accidental global $U(1)$ symmetry because of cross term $\mu \left(
	{\phi^2}^\dagger_1 \phi_2+ \phi^2_1 \phi^\dagger_2 \right)$.  However, in our
	model gauge invariance forbids the inclusion of such cross terms between $\phi_1$
	and $\phi_8$, leading to an accidental global symmetry.
	As a result, after spontaneous symmetry breaking two massless
	Goldstone modes arise such that
	one linear combination of them will be eaten up by the neutral gauge boson corresponding to $U(1)_{B-L}$ gauge group and 
	gives mass to $Z^\prime$ and the other orthogonal combination remains as massless Goldstone 
	boson. We shall  discuss the implications for this massless Goldstone boson in subsequent discussions.

The neutral components of the fields $H$, $\phi_1$ and $\phi_8$ can be parametrised in terms of
real scalars and pseudoscalars as
\begin{align}
&H^0 =\frac{1}{\sqrt{2} }(v+h)+  \frac{i}{\sqrt{2} } A^0\,, \nonumber \\
& \phi_1 = \frac{1}{\sqrt{2} }(v_1+h_1)+  \frac{i}{\sqrt{2} } A_1\,, \nonumber \\
& \phi_8 = \frac{1}{\sqrt{2} }(v_8+h_8)+  \frac{i}{\sqrt{2} } A_8\,.\nn
\end{align}
Here the VEVs of the scalars are given as
 $\langle H\rangle=(0, v/\sqrt2)^T$, $\langle \phi_1\rangle=v_1/\sqrt2$, $\langle \phi_8\rangle=v_8/\sqrt2$.
%================================================ 
%\blue{The minimization conditions imply
%\begin{align}
%\mu^2_H& = -\left(\lambda^2_H v^2 + \frac{ \lambda_{H1}}{2} v^2_1 +   \frac{ \lambda_{H8}}{2} v^2_8 \right),  \nonumber \\
%\mu^2_1& = -\left(\lambda^2_1 v^2_1 + \frac{ \lambda_{H1}}{2} v^2 +   \frac{ \lambda_{18}}{2} v^2_8 \right) , \nonumber \\
%\mu^2_8& = -\left(\lambda^2_8 v^2_8 + \frac{ \lambda_{H8}}{2} v^2 +   \frac{ \lambda_{18}}{2} v^2_1  \right). \label{mincon}
%\end{align}
%}
Then, the CP-even scalar mass matrix can be written as 
\begin{align}
	M_0^2
	=
	\begin{pmatrix}
		2\lambda_H v^2	& \lambda_{H1}vv_1& \lambda_{H8}vv_8	\\
		\lambda_{H1}vv_1	&2\lambda_1 v_1^2	& \lambda_{18}v_1v_8	\\
		\lambda_{H8}vv_8	&\lambda_{18}v_1v_8	& 2\lambda_8 v_8^2				\\
	\end{pmatrix}. 
\end{align}
We consider the Higgs doublet $H$ mixes equally with the two singlets and the mixing is minimal so that the Higgs decay width is consistent with LHC limits. We also assume the VEVs of the singlets $v_1 \simeq v_8 \gg v$ and the couplings $\lambda_1 \simeq \lambda_8$, $\lambda_{H1,H8}\ll \lambda_H$,  then the mass matrix can have the form\footnote{\label{note1}The main point of making  (3.2) and (3.11) in simple form is to give simple analytical expressions for cross section of all the DM annihilation channels. However, the final results are evaluated numerically by scanning over the parameter space of the couplings and hence the exact analytical expressions for the couplings do not affect our main results.}
\begin{align}
	M_0^2
	\simeq
	\begin{pmatrix}
		~a~	& ~a~ & ~a~	\\
		~a~	& ~y~	& ~b~	\\
		~a~	& ~b~	& ~y~ \\
	\end{pmatrix}. 
	\label{massMatrix}
\end{align}
In the limit of minimal Higgs mixing, the unitary matrix connecting flavor and mass eigenstates takes the form
\begin{align}
U \simeq
\begin{pmatrix}
 1 & ~\beta  \cos \alpha -\beta  \sin \alpha  & ~\beta  \cos \alpha +\beta  \sin \alpha  \\
 -\beta  &~ \cos \alpha  & ~\sin \alpha   \\
 -\beta  &~ -\sin \alpha  &~ \cos \alpha  \\
\end{pmatrix},
\end{align}
where $\beta = \frac{a}{b+y-a}$denotes the mixing between $H-\phi_{1,8}$ and  $\alpha = \frac{5\pi}{4}$ is the mixing parameter for $\phi_1-\phi_8$, which are
obtained from the normalized eigenvector matrix  of $M_0^2$ (\ref{massMatrix}). Thus, we obtain  the relation between flavor and mass eigenstates as
\begin{align}
\begin{pmatrix}
 h \\
 h_1 \\
 h_8 \\
 \end{pmatrix} = U \begin{pmatrix} H_1\\ H_2 \\ H_3 \end{pmatrix}=
\begin{pmatrix}
 H_1-H_3 \beta  \sqrt{2} \\
 -H_1 \beta - \frac{H_2}{\sqrt{2}} - \frac{H_3}{\sqrt{2}} \\
 -H_1 \beta + \frac{H_2}{\sqrt{2}} - \frac{H_3}{\sqrt{2}} \\
 \end{pmatrix}.
\end{align}
The various scalar couplings can be expressed as
\begin{eqnarray}
&&2\lambda_{H}v^2 = \lambda_{H1}vv_1 = \lambda_{H8}vv_8 =  \frac{M_{H_1}^2}{(1-2\beta+2\beta^2)},\nn\\
&&2\lambda_{1}v_1^2 = 2\lambda_{8}v_8^2 = \frac{(\beta+1)M_{H_3}^2 + (1+\beta+4\beta^2)M_{H_2}^2}{2(1+\beta+4\beta^2)},\nn\\
&&\lambda_{18}v_1v_8  = \frac{(\beta+1)M_{H_3}^2 - (1+\beta+4\beta^2)M_{H_2}^2}{2(1+\beta+4\beta^2)}.
\label{scouplings}
\end{eqnarray}
Here $H_1$ denotes the SM Higgs with $M_{H_1} = 125.09$ GeV and $v = 246$ GeV.
As discussed earlier, $A_{\rm G}$ appears as the longitudinal polarization of
$Z^\prime$ and the physical massless Goldstone, $A_{\rm NG}$ are given by
\begin{eqnarray}
&&A_{\rm G} = - \frac{8v_8}{\sqrt{v_1^2 + 64v_8^2}}A_8 + \frac{v_1}{\sqrt{v_1^2 + 64v_8^2}}A_1 ,\nn\\
&&A_{\rm NG} = \frac{v_1}{\sqrt{v_1^2 + 64v_8^2}}A_8 + \frac{8v_8}{\sqrt{v_1^2 + 64v_8^2}}A_1.
\end{eqnarray}
As per the assumption $v_1 \simeq v_8$, one can see that $A_{\rm G}$ gets major contribution from $A_8$ and $A_{\rm NG}$ is maximally composed of $A_1$. It should be noted that the massless mode ($A_{\rm NG})$ doesn't couple to any SM particle except Higgs, as we considered non-zero mixing between $H$ and new scalars.
% and it does not violate the cosmological constraints as discussed below. 
It can give rise to an additional decay channel contributing to the invisible width of SM Higgs, given as
\begin{equation}
\Gamma(H_1\to A_{\rm NG}A_{\rm NG}) \simeq \frac{M_{H_1}^3 \sin^2\beta}{32\pi}\left(\frac{v_1^3 + 64v_8^3}{v_1v_8(v_1^2+64v_8^2)}\right)^2~.
\end{equation}
The invisible branching ratio of Higgs is given as
\begin{equation}
\text{Br}_{\rm inv} = \frac{\Gamma(H_1\to A_{\rm NG}A_{\rm NG})}{\Gamma(H_1\to A_{\rm NG}A_{\rm NG}) + \cos^2\beta ~\Gamma^{\rm Higgs}_{\rm SM}}.
\end{equation}
Using the constraint, $\text{Br}_{\rm inv}\simeq 20\%$ \cite{Belanger:2013kya,Giardino:2013bma}, $\Gamma^{\rm Higgs}_{\rm SM} \simeq 4$ MeV, we obtain the upper limit on the mixing angle as
\begin{equation}
|{\tan\beta}| \lesssim 2.2\times 10^{-4}\times \left(\frac{v_1}{\text{GeV}}\right).
\end{equation}
Moreover, if the NG stays in thermal equilibrium with ordinary matter until muon annihilation, then it mimics as fractional cosmic neutrinos contributing nearly 0.39 to the effective number of neutrino species \cite{Weinberg:2013kea, Garcia-Cely:2013nin} to give $N_{\rm eff} = 3.36^{+0.68}_{-0.64}$ at $95\%$ C.L, a remarkable agreement with Planck data \cite{Ade:2013zuv}. 
%It was first explained in \cite{Weinberg:2013kea} taking $\lambda_{1} \simeq 0.005$ and $M_{H_2} \simeq 500$ MeV. 
This illustration was done by working in the low mass regime of the physical scalar ($\simeq 500$ MeV) \cite{Weinberg:2013kea}. However, in \cite{Garcia-Cely:2013nin} it was found that for masses $\gtrsim$ 4 GeV the Goldstone bosons do not contribute to $N_{\rm eff}$.
And since in the present work we consider higher mass regime for the physical scalar spectrum to discuss the effect of NG on relic density, the contribution of NG to $N_{\rm eff}$ is not applicable.
\subsection{Mixing in fermion sector}
%================================================
The heavy Majorana mass matrix is given by
%================================================
\begin{align}
	M_R
	=
	\begin{pmatrix}
		y_{11} \langle \phi_8 \rangle	& y_{12} \langle \phi_8 \rangle	& y_{13} \langle \phi_1 \rangle	\\
		y_{12} \langle \phi_8 \rangle	& y_{22} \langle \phi_8 \rangle	& y_{23} \langle \phi_1 \rangle	\\
		y_{13} \langle \phi_1 \rangle	& y_{23} \langle \phi_1 \rangle	& 0				\\
	\end{pmatrix} \;.
	\label{massMatrix1}
\end{align}
%================================================
%================================================
%The masses of other SM gauge bosons are coming from the VEV of the SM Higgs boson. Assuming no kinetic mixing among the $U(1)$ gauge bosons, 
%the resulting neutral gauge boson mass matrix  in the basis $(W_{\mu}^{3},~B_{\mu},~Z_\mu^\prime)$
%reads as
%%================================================
%\begin{eqnarray}
%	\mathcal{M}_\text{neutral}^2
%	=
%	\left(\begin{array}{ccc}
%		\frac{1}{8} g^2 v^2	& -\frac{1}{8} g g^\prime v^2			& 0	\\
%		-\frac{1}{8} g g^\prime v^2	& \frac{1}{8} g^{\prime^2} v^2	& 0	\\
%		%
%		0	& 0	& g^2_{BL} \left(  v_1^2+ 64  v_8^2 \right)									\\
%\end{array} \right)
%\, .
%\end{eqnarray}
%%
%Diagonalization of the above matrix gives  the  physical mass of the extra neutral gauge boson $Z^\prime$ as
%%================================================
%\begin{equation}
%	M^2_{Z^\prime}
%	=
%	 g^2_{BL} \left(v_1^2+ 64 v_8^2 \right),
%	\end{equation}
%the SM $Z$ mass becomes $M_Z^2=\frac{1}{4} v^2 (g^2+g^{\prime^2})$ and the photon remains massless.  
For simplicity, we  consider the above mass matrix with real entries of the form\textsuperscript{\ref{note1}}
\begin{align}
	M_R
	=
	\begin{pmatrix}
		x~	&~ a~& ~b	\\
		a~	&~ x~	& ~b	\\
		b~	&~ b~	& ~0				\\
	\end{pmatrix}, 
	%\label{massMatrix}
\end{align}
which can be obtained by assuming  the Yukawa couplings to satisfy the relations $y_{11}\approx y_{22}$ and $y_{13} \approx y_{23}$ along with $v_1 \approx v_8$.
The above mass matrix can be diagonalized using the unitary matrix  as $(U_1\cdot K)^T \cdot M_R \cdot (U_1\cdot K)$, where  $U_1$ is the normalized eigenvector matrix  of $M_R$ and
% \begin{align}
%U_1 =  \begin{pmatrix}
%		-\frac{1}{\sqrt{2}} & -\frac{1}{\sqrt{6}}	& ~\frac{1}{\sqrt{3}}	\\
%		\frac{1}{\sqrt{2}}	&   -\frac{1}{\sqrt{6}} & ~\frac{1}{\sqrt{3}}	\\
%		0	& \frac{\sqrt{2}}{\sqrt{3}}	& ~\frac{1}{\sqrt{3}}				\\
%	\end{pmatrix}.
%\end{align}
$K = {\rm{diag}}(1,i,1)$ is a diagonal phase matrix used to avoid the negative mass eigenvalues. Thus, one obtains the mass matrix in the diagonal basis $M^{\rm diag}= {\rm{diag}}(M_{D1},M_{D2},M_{D3})$ as
\begin{align}
 M^{\rm diag}=  \begin{pmatrix}
		x-a	& 0	& 0	\\
		0	& \frac{1}{2}\left(-(x+a) + \sqrt{8b^2 + (x+a)^2}\right)	& 0	\\
		0	& 0	& \frac{1}{2}\left((x+a) + \sqrt{8b^2 + (x+a)^2}\right)\\
	\end{pmatrix}.
	%\label{massMatrix}
\end{align}
To make the analysis simpler, we consider $M_{D2} = \frac{1}{2}M_{D3}$, which implies $b = x+a$. Thus, the final diagonal matrix\footnote{The lightest mass eigenstate is taken as the dark matter candidate while the heavier ones are taken to be sufficiently massive such that they decouple from the phenomenology and play no role in our final results.  To illustrate this scenario, we make the assumption of $M_{D2} = \frac{M_{D3}}{2}$ which makes $N_{D1}$ the dark matter candidate while $N_{D2}$ and $N_{D3}$ are very massive and effectively decouple from the theory.} is given by
\begin{align}
{\rm{diag}}(M_{D1},M_{D2},M_{D3}) =  \begin{pmatrix}
		x-a	& 0	& 0	\\
		0	& x+a	& 0	\\
		0	& 0	& 2(x+a)				\\
	\end{pmatrix}.
	\label{mass evalues}
\end{align}
Considering $x > a$, we get positive eigenvalues and the mass eigenstates $N_{Di}$ can be written as
\begin{eqnarray}
N_{D1} &=& \frac{N_{\text{2}}-N_{\text{1}}}{\sqrt{2}}, \nn\\
N_{D2} &=& \frac{i \left(N_{\text{1}}+N_{\text{2}}-2 N_{\text{3}}\right)}{\sqrt{6}}, \nn\\
N_{D3} &=& \frac{N_{\text{1}}+N_{\text{2}}+N_{\text{3}}}{\sqrt{3}}.\label{mass estates}
\end{eqnarray}
The Yukawa couplings can be expressed in terms of the physical masses as
\begin{eqnarray}
y_{11}&=&y_{22}=  \frac{\sqrt{2}~( M_{D1}+M_{D2})}{2 v_8}\;,\nonumber\\
y_{12}&=& \frac{\sqrt{2}~(-M_{D1}+M_{D2})}{2 v_8}\;,\nonumber\\
y_{13}&=&y_{23}= \frac{\sqrt{2}~M_{D2}}{ v_1}\;.
\label{yukf}
\end{eqnarray}
The interaction terms between the new fermions and the $Z'$ gauge boson can be written in the mass eigenstate basis as 
\begin{eqnarray}
{\cal L}^V_{N_{Di}} &=& g_{BL} \biggr [-4 \overline{N^c_{D1}} \gamma^\mu N_{D1} + 2\overline{N^c_{D2}} \gamma^\mu N_{D2} - \overline{N^c_{D3}} \gamma^\mu N_{D3} \nn\\ &-& 3i \sqrt{2}~ \overline{N^c_{D2}} \gamma^\mu N_{D3} + 3i \sqrt{2}~\overline{N^c_{D3}} \gamma^\mu N_{D2}  \biggr] Z'_\mu\;. 
\end{eqnarray}
Similarly, the interaction terms with the singlets $\phi_1$ and $\phi_8$ are 
\begin{eqnarray}
{\cal L}^S_{N_{Di}} &=&   (y_{11}-y_{12})\overline{N^c_{D1}}\,N_{D1} \,\phi_8   +   \frac{1}{3}\left (  4 y_{13} \overline{N^c_{D2}}\,N_{D2} \phi_1 -(y_{11}+y_{12}) \overline{N^c_{D2}}\,N_{D2} \phi_8  \right) \nn\\
&+&  \frac{2}{3} \left(2 y_{13} \overline{N^c_{D3}}\,N_{D3} \phi_1 + (y_{11}+y_{12}) \overline{N^c_{D3}}\,N_{D3} \phi_8 \right) \nn\\
&+& \frac{2\sqrt{2} i}{3} \left( y_{13}\overline{N^c_{D2}}\,N_{D3} \phi_1 -(y_{11} + y_{12})\overline{N^c_{D2}}\,N_{D3} \phi_8  \right).\label{scalar1}
\end{eqnarray}
A glance at Eqns. (\ref{mass evalues}), and (\ref{mass estates}) confirms that $N_{D1}$ is the lightest Majorana mass eigenstate and we intend to perform a detailed study of Majorana dark matter in this work.
\section{Dark Matter phenomenology}
Since the proposed dark matter particle $N_{D1}$, the lightest of
the three Majorana states, which  can interact with the scalar sector and vector gauge boson $Z^{\prime}$, the model can be well explored in dark matter observables in this dual portal scenarios separately\footnote{
The phenomenological study is quite different as we shall see that WIMP-nucleon cross-section is insensitive to direct detection experiments in $Z^{\prime}$-portal, whereas one can have stringent experimental limits in the scalar-portal. Moreover, the discussion becomes more transparent as the limits from ATLAS and LEP-II are only applicable in $Z^{\prime}$-mediated observables and the effect of massless Goldstone is visible only in scalar mediated DM relic density.}. In this section, we will be discussing the DM phenomenology in our model.  
%The lightest of
%the three Majorana states represents the dark matter candidate which can interact with the
%SM particles through the $Z^\prime$ boson as well as the additional scalars in the scalar
%sector.  
We begin our discussion with relic density constraints on Majorana dark matter in the $B-L$ model
considered here.
\subsection{Relic density for Majorana dark matter}
We first present the analytical expressions for annihilation cross sections that contribute to relic
density in our model.

The formula used for computing the relic abundance of dark matter is  
\begin{equation}
\Omega \text{h}^2 = \frac{2.14 \times 10^{9} ~{\rm{GeV}}^{-1}}{ {g_\ast}^{1/2} M_{\rm{pl}}}\frac{1}{J(x_f)}\;,
\label{eq:relicdensity}
\end{equation}
where the Planck mass $M_{\rm{pl}}=1.22 \times 10^{19} ~\rm{GeV}$, $g_\ast = 106.75$ being the total
number of effective relativistic degrees of freedom, and $J(x_f)$ reads as
\begin{equation}
J(x_f)=\int_{x_f}^{\infty}  \frac{dx}{x^2} \langle \sigma v \rangle (x).
\end{equation}
The freeze out parameter $x_f$ in the above integral is given as
\begin{equation}
x_f= \ln \left( \frac{0.038 \ g \ M_\text{Pl} \ M_{D1} \ \langle\sigma v\rangle (x_f) }{({g_\ast
x_f})^{1/2}} \right),
\end{equation}
where $g$ is the count of number of degrees of freedom of the dark matter particle. The thermally
averaged annihilation cross section $\langle \sigma v \rangle$ is given by
\begin{equation}
\langle\sigma v\rangle (x) = \frac{x}{8 M_{D1}^5 K_2^2(x)} \int_{4 M_{D1}^2}^\infty ds\:
\hat{\sigma} \times ( s - 4 M_{D1}^2) \ \sqrt{s} \ K_1 \left(\frac{x \sqrt{s}}{M_{D1}}\right),
\end{equation}
where $K_1$, $K_2$ denote the modified Bessel functions and $x = M_{D1}/T$, with $T$ being the temperature.  
%  Once the masses of the physical scalars are defined, the couplings in Eqn. (\ref{scoup}) are known. 
%The parameters that are fixed during the analysis are given in Table \ref{fixedh}.  
We now discuss the different annihilation channels that contribute to the relic density and the
impact of the different parameters in both scalar and vector mediated DM scenarios.
\subsubsection{Scalar mediated}

\begin{figure}[thb]
\centering
%\begin{center}
\subfloat[]{\includegraphics[width=0.3\linewidth]{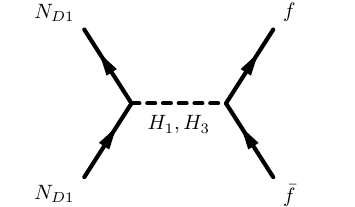}}
%\vspace{0.01 cm}
\subfloat[]{\includegraphics[width=0.3\linewidth]{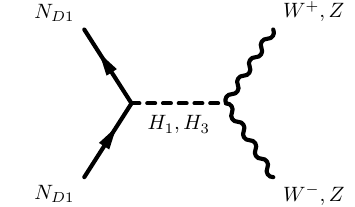}}
%\vspace{0.01 cm}
\subfloat[]{\includegraphics[width=0.3\linewidth]{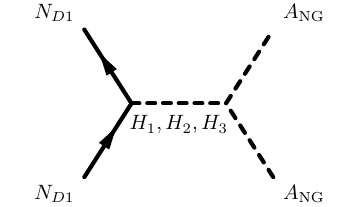}}
\vspace{0.01 cm}
\subfloat[]{\includegraphics[width=0.3\linewidth]{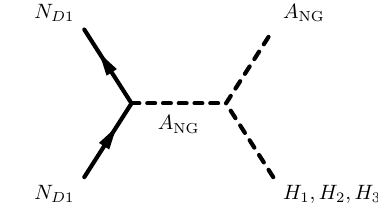}}
\vspace{0.01 cm}
\subfloat[]{\includegraphics[width=0.3\linewidth]{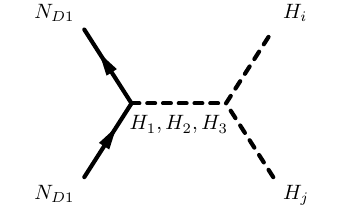}}
\vspace{0.01 cm}
\subfloat[]{\includegraphics[width=0.3\linewidth]{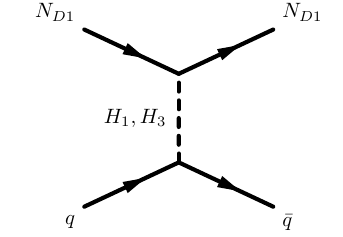}}
\caption{Feynman diagrams contributing to relic density are shown in Figs. (a) - (e) while Fig.  (f) is relevant for the direct searches.}
\label{hfeyn}
%\end{center}
\end{figure}
\begin{figure}[t!]
\begin{center}
\includegraphics[width=0.48\linewidth]{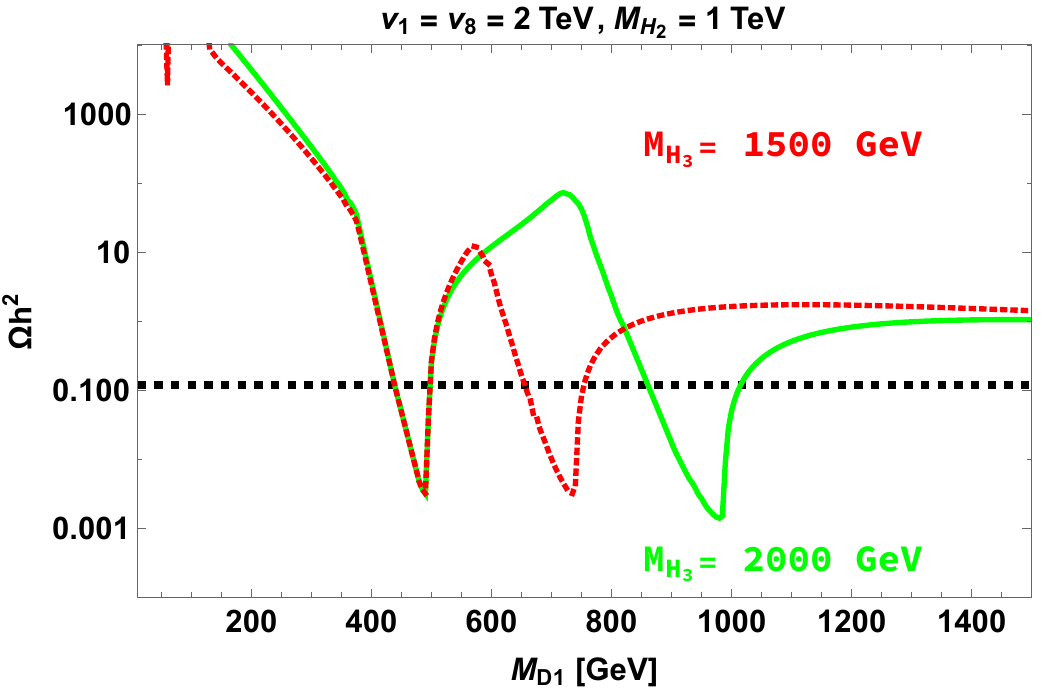}
\caption{Scalar-portal relic abundance as a function of DM mass $M_{D1}$ for two specific mass values of the physical scalar $H_3$. The horizontal dashed lines represent the $3\sigma$ value of the current relic density \citep{Ade:2015xua}.}
\label{omegaH}
\end{center}
\end{figure}

The possible annihilation channels that can drive the relic density in the scalar portal scenario
are shown in the first five Feynman diagrams of Fig. \ref{hfeyn}.   These channels can be either SM fermions,
SM gauge bosons ($W,Z$), Higgs sector scalars and the massless physical Goldstone mode.  The cross section for the annihilation channels into SM fermions and gauge bosons are given by
\begin{align}
	\hat{\sigma}^S_{ff} =& \frac{C}{v^2s}|F_1|^2
\sum_f M^2_f c_f (s-4 M^2_{f})(s - 4M^2_{D1}) \frac{(s-4 M^2_{f})^\frac{1}{2}}{(s - 4M^2_{D1})^{\frac{1}{2}}}\;,\\
\hat{\sigma}^S_{WW} =& \frac{Cs}{2v^2}|F_1|^2
(s - 4M^2_{D1})\left(1- \frac{4 M^2_W}{s} + \frac{12 M^4_W}{s^2}\right) \frac{(s-4
	M^2_{W})^\frac{1}{2}}{(s - 4M^2_{D1})^{\frac{1}{2}}}\;,\\
\hat{\sigma}^S_{ZZ} =&\frac{Cs}{4v^2}|F_1|^2  (s - 4M^2_{D1})\left(1- \frac{4 M^2_Z}{s} + \frac{12
M^4_Z}{s^2}\right) \frac{(s-4 M^2_{Z})^\frac{1}{2}}{(s - 4M^2_{D1})^{\frac{1}{2}}}\;,
\end{align}
while the expressions for channels with NG in final state turn out to be
\begin{align}
	\hat{\sigma}^S_{\rm{NG}} =&\frac{Cs}{4 v_1^2 v_8^2(v_1^2 + 64v_8^2)^2}|F_2|^2 
(s - 4M^2_{D1}) \frac{s^\frac{1}{2}}{(s - 4M^2_{D1})^{\frac{1}{2}}}\;,\\
	\hat{\sigma}^S_{\rm{NG~H_1}} =&\frac{C \beta^2}{2v_8^2(v_1^2 + 64v_8^2)^3} \left(v_1^3 + 64 v_8^3\right)^2
\frac{(s - 4M^2_{D1})^\frac{1}{2}(s-M_{H_1}^2)^3 }{s^\frac{7}{2}}\;,\\
	\hat{\sigma}^S_{\rm{NG~H_2}} =&\frac{C  }{4v_8^2(v_1^2 + 64v_8^2)^3} \left(v_1^3 - 64 v_8^3\right)^2
\frac{(s - 4M^2_{D1})^\frac{1}{2}(s-M_{H_2}^2)^3 }{s^\frac{7}{2}}\;,\\
	\hat{\sigma}^S_{\rm{NG~H_3}} =&\frac{C }{4v_8^2(v_1^2 + 64v_8^2)^3} \left(v_1^3 + 64 v_8^3\right)^2
\frac{(s - 4M^2_{D1})^\frac{1}{2}(s-M_{H_3}^2)^3 }{s^\frac{7}{2}}\;,
\end{align}
where
\begin{align}
	C=&\frac{ (y_{11}-y_{12})^2}  { 8\pi }\;,\\
	F_1=&-\frac{\beta}{\left[(s-M^2_{H_1}) + i M_{H_1} \Gamma_{H_1}\right]}
+ \frac{\beta}{\left[(s-M^2_{H_3}) + i M_{H_3} \Gamma_{H_3}\right]}\;,\\
F_2=&\frac{\beta^2 \left(v_1^3 + 64 v_8^3\right)}{\left[(s-M^2_{H_1}) + i M_{H_1} \Gamma_{H_1}\right]} +
	\frac{1/2\left(v_1^3 - 64 v_8^3\right)}{\left[(s-M^2_{H_2}) + i M_{H_2} \Gamma_{H_2}\right]} + 	\frac{1/2\left(v_1^3 + 64 v_8^3\right)}{\left[(s-M^2_{H_3}) + i M_{H_3} \Gamma_{H_3}\right]}\;,
\end{align}
with $c_f$ and $M_f$ denoting the color charge and mass of the the SM fermion $f$ respectively. 
Finally, the Higgs sector annihilation channels  we have
\begin{equation}
	\hat{\sigma}^S_{H_i H_j} = \frac{C}{2s~n!}|F_{ij}|^2 
(s - 4M^2_{D1}) \frac{\left [(s- (M_{H_i}+ M_{H_j})^2)(s- (M_{H_i}- M_{H_j})^2) \right ]^\frac{1}{2}}{\left [s(s - 4M^2_{D1})\right ]^{\frac{1}{2}}}\;,
\label{hann}
\end{equation}
where
\begin{align*}
F_{ij}=&-\frac{\lambda_{1ij}\beta}{\left[(s-M^2_{H_1}) + i M_{H_1} \Gamma_{H_1}\right]} +
	\frac{\lambda_{2ij}/\sqrt{2}}{\left[(s-M^2_{H_2}) + i M_{H_2} \Gamma_{H_2}\right]} - 	\frac{\lambda_{3ij}/\sqrt{2}}{\left[(s-M^2_{H_3}) + i M_{H_3} \Gamma_{H_3}\right]} \;,
\end{align*}
where $n$ denotes the permutation factor for identical final state particles and $\lambda_{1ij}, \lambda_{2ij}, \lambda_{3ij}$ having mass dimension denote the trilinear scalar couplings with $i,j = 1,2,3$. In the analysis, we consider the VEVs $v_1$ and $v_8$ to be in TeV scale range so that the scalar couplings (\ref{scouplings}) are within the limits of unitarity bounds (\ref{unitarity}). The mixing parameter $\beta$ can be written in terms of the physical scalar masses as
\begin{equation}
\beta=\frac{-M_{H_1}^2+M_{H_3}^2-\sqrt{-15 M_{H_1}^4-10 M_{H_3}^2 M_{H_1}^2+M_{H_3}^4}}{4 \left(2 M_{H_1}^2+M_{H_3}^2\right)}.
\end{equation}
Since the Higgs mass ($M_{H_1}$) is fixed, the mass parameter $M_{H_3}$ defines the amount of mixing i.e., say $M_{H_3} \ge 1$ TeV implies $\beta \le 0.016$.
Fig. \ref{omegaH} displays the behavior of relic density with the dark matter mass where the PLANCK limit is reached on the either side of resonance of the propagators. For lower DM mass region, the channels $f\bar{f}$ and $A_{\rm NG} A_{\rm NG}$ maximally contribute to relic density. Then, the rest of channels contribute to relic density once they get kinematically allowed. The channels with $H_1H_1$ and $A_{\rm NG}A_{\rm NG}$ in final state can give the resonance in $H_2$ propagator. Emphasis is given more to the mass of $H_3$ as the WIMP-nucleon cross section also involves this mass parameter.
%
%
%\begin{equation}
%\Gamma_{H_1} = \left(\frac{M_f {\rm{cos}}\alpha}{v}\right)^2 \frac{M_{H_1}}{8 \pi} \left(1-\frac{4 M_f^2}{M_{H_1}^2} \right)^{3/2},
%\end{equation}
%\begin{equation}
%\Gamma_{H_2} = \left(\frac{M_f {\rm{sin}}\alpha}{v}\right)^2 \frac{M_{H_2}}{8 \pi} \left(1-\frac{4 M_f^2}{M_{H_2}^2} \right)^{3/2}.
%\end{equation}
%\begin{table}[t!]
%\begin{center}
%================================================
%\begin{tabular}{|c|c|c|c|c|c|c|c|c|c|}
%	\hline
%	Parameters 		&   $M_{h_{1}}$ [GeV] & $v_1$ [GeV] \\
%	\hline
%	Values &  &  $1000$ & $1000$ \\
%	\hline
%	\hline
%\end{tabular}
%\caption{Fixed parameters for $Z^{\prime}$-mediated DM observables}
%\label{fixexzp}
%\end{center}
%\end{table}
%================================================
%
%
%\begin{table}[htb]
%\begin{center}
%%================================================
%\begin{tabular}{|c|c|c|}
%	\hline
%	Parameters & Scalar portal & Vector portal \\\hline
%	$M_{H_{1}}$ [GeV] & $125.09$ & $125.09$ \\
%	$M_{H_{2}}$ [GeV] & $1000-2000$ & $1000$ \\
%	$M_{H_{3}}$ [GeV] & $M_{H_2}-3000$ & $1000$ \\
%	$v$ [GeV] & $246$ & $246$ \\
%	$v_1$ [GeV] & $1000-3000$ & $1000$ \\
%	$v_8$ [GeV] & $1000-3000$ & $1000$ \\
%	\hline
%	\hline
%\end{tabular}
%\caption{Fixed parameters for scalar and vector mediated DM observables.}
%\label{fixedh}
%\end{center}
%\end{table}
\subsubsection{Vector mediated}
\begin{figure}[t!]
\begin{center}
\includegraphics[width=0.3\linewidth]{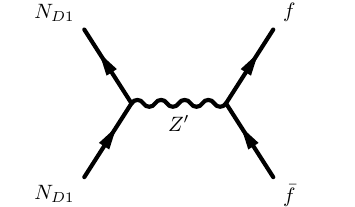}
\vspace{0.01 cm}
\includegraphics[width=0.3\linewidth]{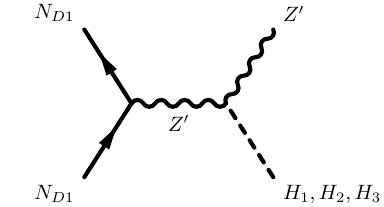}
\vspace{0.01 cm}
\caption{Feynman diagrams contributing to relic density in the vector-mediated case.}
\label{zpfeyn}
\end{center}
\end{figure}

The DM also interacts with the visible sector through the gauge mediated processes which can lead to
annihilation channels into SM fermions and the Higgs sector as shown in Fig. \ref{zpfeyn}.  The cross sections are given by 
\begin{align}
\hat{\sigma}^V_{ff} =& \sum_f \frac{  16 (n^f_{\rm BL})^2g^4_{\rm BL} c_f |F_V|^2}  {3 \pi s } (s-4 M^2_{ D1})(s + 2M^2_f) \frac{(s-4 M^2_f)^{\frac{1}{2}}}{(s-4 M^2_{ D1})^{\frac{1}{2}}}\;,\nn\\
\hat{\sigma}^V_{Z^{\prime} H_3} =& \frac{4(64v_8 + v_1)^2 g_{BL}^6  |F_V|^2}{\pi s} \frac{((s-
	(M_{Z^{\prime}}+ M_{H_3})^2)(s- (M_{Z^{\prime}}- M_{H_3})^2))^\frac{1}{2}}{(s(s -
		4M^2_{D1}))^{\frac{1}{2}}}C_{H_3}\;,\nn\\
\hat{\sigma}^V_{Z^{\prime} H_2} =& \frac{4(64v_8 - v_1)^2 g_{BL}^6  |F_V|^2}{\pi s} \frac{((s-
	(M_{Z^{\prime}}+ M_{H_2})^2)(s- (M_{Z^{\prime}}- M_{H_2})^2))^\frac{1}{2}}{(s(s -
		4M^2_{D1}))^{\frac{1}{2}}}C_{H_2}\;,\nn\\
\hat{\sigma}^V_{Z^{\prime} H_1} =& \frac{8(64v_8 + v_1)^2 \beta^2 g_{BL}^6  |F_V|^2}{\pi s} \frac{((s-
	(M_{Z^{\prime}}+ M_{H_1})^2)(s- (M_{Z^{\prime}}- M_{H_1})^2))^\frac{1}{2}}{(s(s -
		4M^2_{D1}))^{\frac{1}{2}}}C_{H_1}\;,
\end{align}
where
\begin{align*}
	F_V=&\frac{1}{\left[(s-M^2_{Z^{\prime}}) + i M_{Z^{\prime}} \Gamma_{Z^{\prime}}\right]}\;,\\
    C_S =& \left[ \frac{(s-8M_{D1}^2)}{4} + \frac{1}{M_{Z^{\prime}}^2  } \left(2 s M_{D1}^2 + \frac{(s
	    + M_{Z^{\prime}}^2 - M_{S}^2)^2}{4}\right.\right.\\ 
    &\left.\left.-\frac{1}{48s} (s-4M_{D1}^2)(s-(M_{Z^{\prime}}+M_{S})^2)(s-(M_{Z^{\prime}}-M_{S})^2)\right) \right],
	\label{}
\end{align*}
%\begin{align*}
%	C_S^{\prime} =&  256 v_8 \cos \alpha,\\
%	=& 256 v_8 \sin \alpha,\\
%	=& 4 v_1.
%	\label{}
%\end{align*}
\begin{figure}[t!]
\begin{center}
\includegraphics[width=0.48\linewidth]{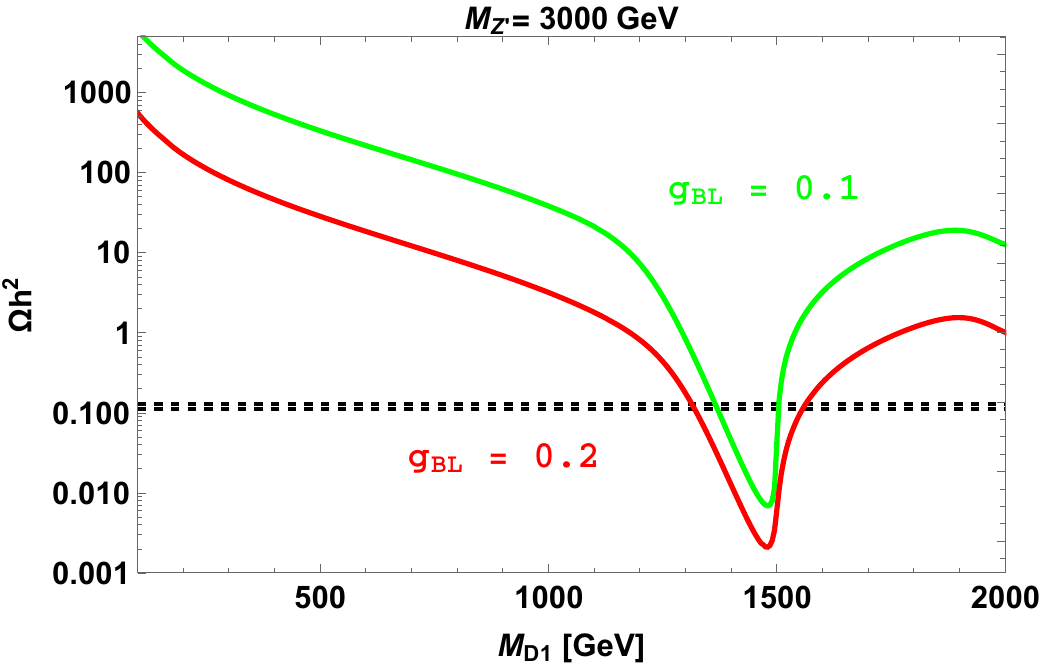}
\vspace{0.2 cm}
\includegraphics[width=0.48\linewidth]{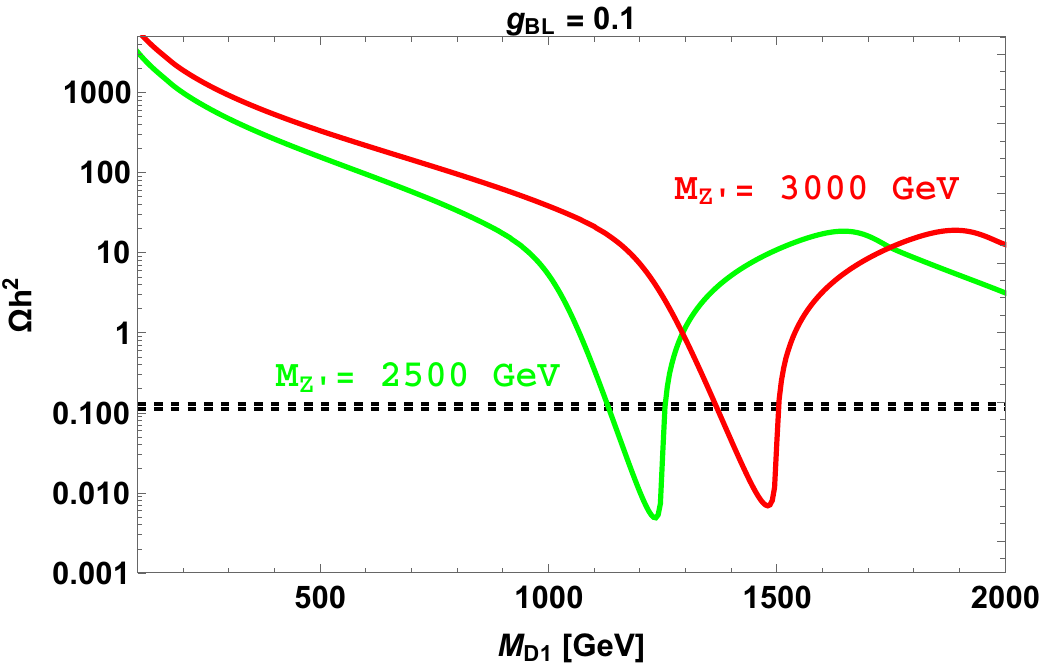}
\caption{Variation of relic abundance $\Omega \text{h}^2$ with the mass of DM with  $(M_{H_2},M_{H_3}) = (1,1.5)$ TeV. Left panel depicts the variation for fixed $Z^{\prime}$ mass and varying $B-L$ gauge coupling $g_{BL}$. The right panel displays the behavior for constant coupling $g_{BL}$ and varying mediator mass. Here, the horizontal dashed lines represent the $3\sigma$ value of the current relic density \citep{Ade:2015xua}.}
\label{omegazp}
\end{center}
\end{figure}
with $S=H_1,H_2,H_3$.
Here $n^f_{\rm BL}$ denotes the $B-L$ charge for the SM fermion $f$ and
$\Gamma_{Z^{\prime}}$ is the decay width of the heavy gauge mediator $Z^{\prime}$.  We use the packages LanHEP \cite{Semenov:1996es}, micrOMEGAs \cite{Pukhov:1999gg,
Belanger:2006is, Belanger:2008sj} to compute the DM observables. Fig.
\ref{omegazp} shows the behaviour of relic abundance with the mass of dark matter particle
for various sets of gauge coupling $g_{BL}$ and the mediator mass $M_{Z^{\prime}}$
consistent with the LEP-II bound \citep{Schael:2013ita} i.e., $M_{Z^\prime}/g_{BL}>7$ TeV. Near the resonance the major contribution comes from the $N_{D1} N_{D1}
\rightarrow f\bar{f}$ channel. As we go towards high mass regime of $M_{D1}$, the channels
$N_{D1} N_{D1} \rightarrow Z^{\prime} H_{1,2,3}$ become
dominant resulting in a slight decrease in the relic abundance.
\subsection{Direct searches}
%
%\subsection{Scalar mediated}
In this section, we discuss the direct detection prospects for our model in both scalar and
vector mediated DM scenarios.  
Since the vector boson $Z^{\prime}$ couples differently to Majorana fermion and quarks
i.e., axial vector and vector type, the contribution by WIMP-nucleon interaction is
insensitive to direct detection experiments
\cite{Agrawal:2010fh,Zheng:2010js,Okada:2012sg}.  Hence, we shall only focus on the scalar
mediated DM scattering and constraints on it from various experiments.
The effective Lagrangian term of scalar mediated channel shown in  panel of Fig. \ref{hfeyn}-(f) that contributes to the
spin-independent (SI) cross section for direct detection is
\begin{equation}
\mathcal{L}_{\mathrm{eff}}= a_q  \overline{N}_{D1} N_{D1} \bar{q} q,
\end{equation}
where \begin{equation}
 \frac{a_q}{M_q} =  \frac{ (y_{11}-y_{12})\beta}{\sqrt{2}v} \left(
 \frac{1}{M^2_{H_3}} - \frac{1}{M^2_{H_1}} \right).
 \label{aq}
\end{equation}
The WIMP-nucleon SI contribution reads as
\begin{equation}
\sigma_{\rm SI} = \frac{4}{\pi}
\left(\frac{M_p M_{D1}}{M_p + M_{D1}}\right)^2 f_p^2\;,%\left[ f_p Z+f_n (A-Z)\right]^2\;, 
 \label{sigmaSI}
\end{equation}
where $M_p$ denotes the mass of proton and the hadronic matrix element $f_p$ is given as
\begin{equation}
 \frac{f_p}{M_p} = \sum_{q=u,d,s}f_{Tq}^{p}\frac{a_q}{M_q} 
  + \frac{2}{27}\left(1-\sum_{q=u,d,s}f_{Tq}^{p}\right)\sum_{q=c,b,t}\frac{a_q}{M_q}\;.
\end{equation}
Typical values for proton are $f_{Tu}^{p} = 0.020\pm 0.004$,  $f_{Td}^{p} = 0.026\pm0.005$
and $f_{Ts}^{p} = 0.118\pm0.062$ \cite{Ellis:2000ds}. 
% Here we use micrOMEGAs to calculatethe form factors which depend on the nucleon sigma terms $\sigma_{\pi N}$ and
%$\sigma_s$ (see \cite{Alarcon:2011zs,Alarcon:2012nr} for estimates).
%
\begin{figure}[t!]
\begin{center}
\includegraphics[width=0.48\linewidth]{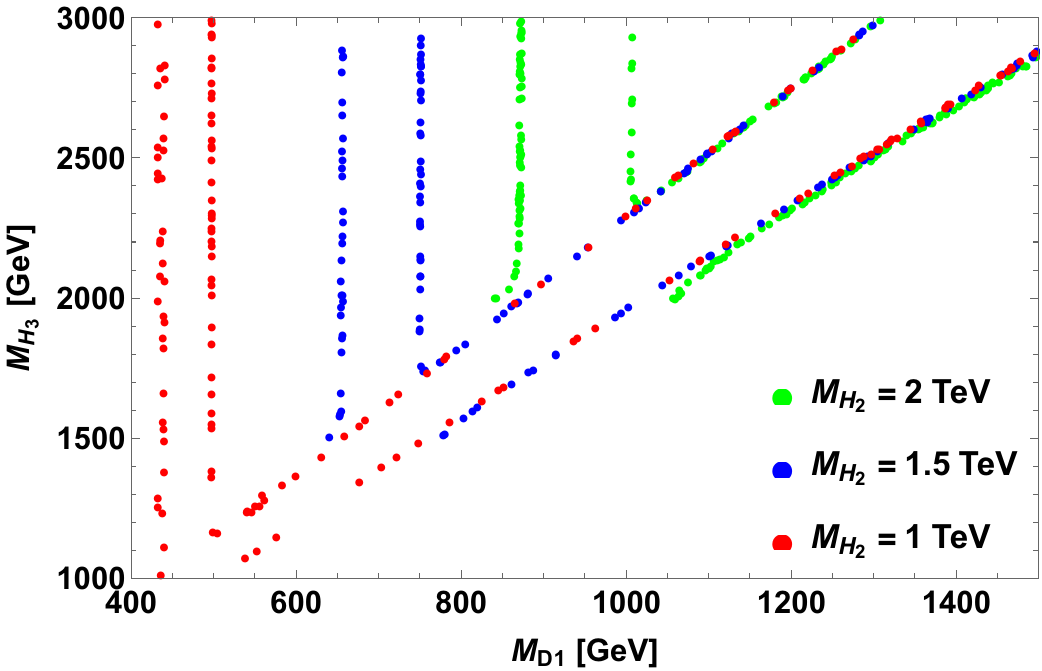}
\vspace{0.2 cm}
\includegraphics[width=0.48\linewidth]{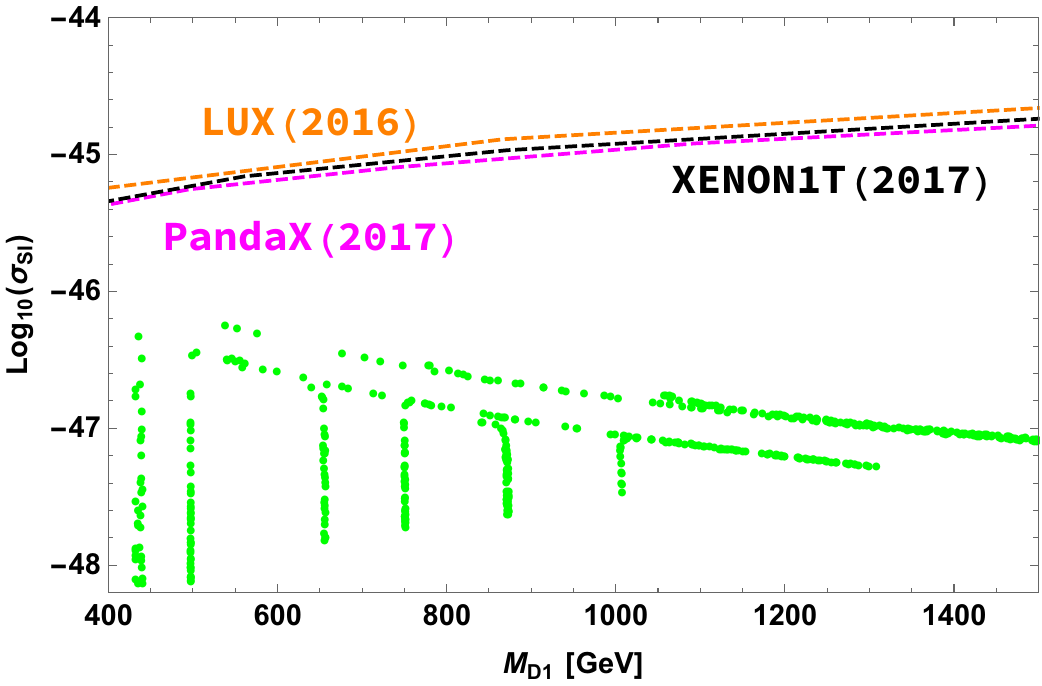}
\caption{Left panel shows the parameter space satisfying the $3\sigma$ range in current relic density and the most stringent PandaX limit. Right panel depicts WIMP-nucleon cross section for the parameters space depicted in the left panel. The dashed lines denote the upper bound on SI cross section from LUX \cite{Akerib:2016vxi}, XENON1T    \cite{Aprile:2017iyp} and PandaX \cite{Cui:2017nnn}.}.
\label{DD}
\end{center}
\end{figure}
\begin{table}[htb]
\begin{center}
%================================================
\begin{tabular}{|c|c|}
	\hline
	Parameters & Range  \\\hline
	$v_{1,8}$ [GeV] & $2000$\\
	$M_{H_{2}}$ [GeV]  & $1000-2000$\\
	$M_{H_{3}}$ [GeV]  & $M_{H_2}-3000$\\
	$\beta$ & $0.016-0.0016$ \\
	\hline
	\hline
\end{tabular}
\caption{Parameters and their ranges for scalar portal analysis.}
\label{paraH}
\end{center}
\end{table}
Varying the parameters in the range shown in Table. \ref{paraH}, we show in Fig. \ref{DD} (left panel), the parameter space that satisfies the $3 \sigma$ range in the
current relic density \cite{Ade:2015xua} and the PandaX limit \cite{Cui:2017nnn}. Since the mixing parameter $\beta$ is small, the direct detection limits on the parameter space is not stringent. It is mainly constrained by relic density where the PLANCK limit is met near the resonance in two propagators $H_2$ (vertical data points) and $H_3$ (diagonal data points). Right panel depicts the WIMP-nucleon cross section with varying mass of the DM of the parameter space shown in the left panel.
\section{Collider studies}
In recent past, both ATLAS and CMS experiments have provided extensive studies to search for new
heavy resonances in both dilepton and dijet signals. 
It is found that these two experiments provide lower limit on ${Z^{\prime}}$-boson with dileptons,
resulting in stronger bounds than dijets due to relatively 
fewer background events.  ATLAS results \cite{ATLAS-CONF-2015-070} from the study of dilepton signals for the $Z^{\prime}$
boson provide the most stringent limits on the heavy gauge boson mass $M_{Z^{\prime}}$ 
and the gauge coupling $g_{BL}$. 

\begin{figure}[t!]
\begin{center}
\includegraphics[width=0.48\linewidth]{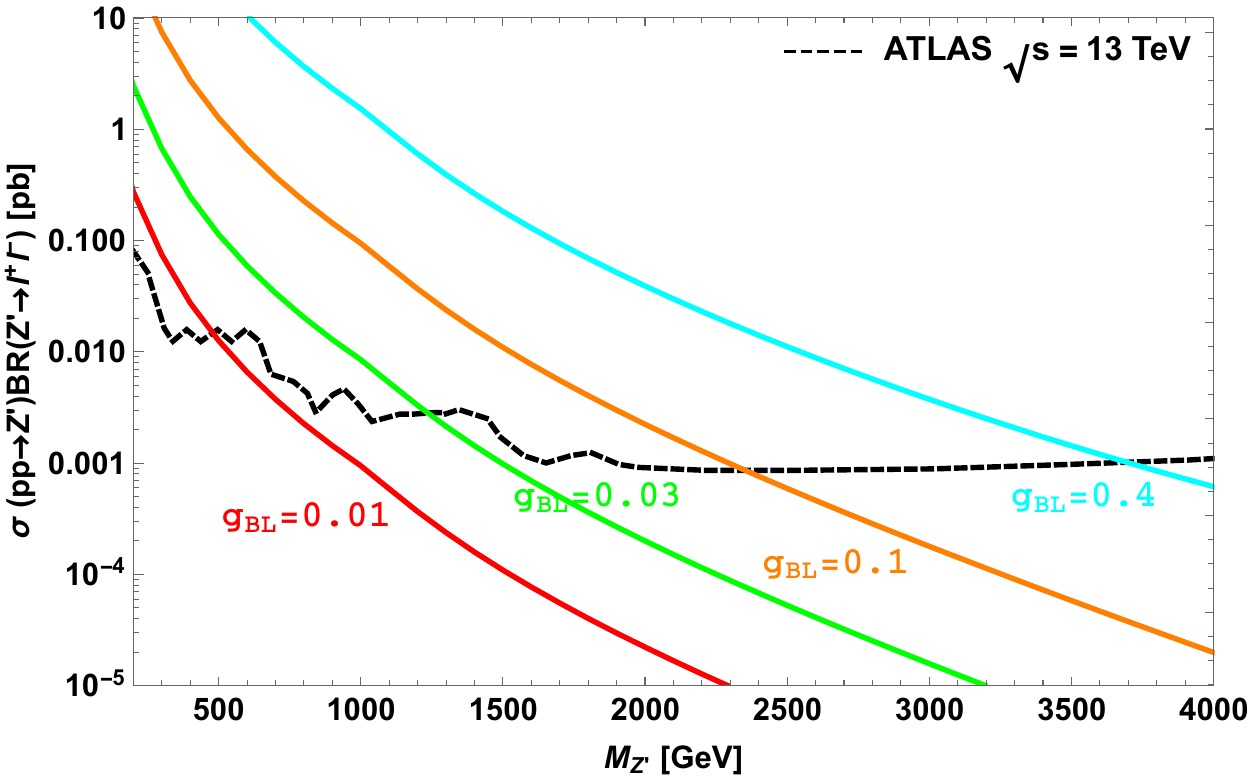}
\vspace{0.2 cm}
\includegraphics[width=0.48\linewidth]{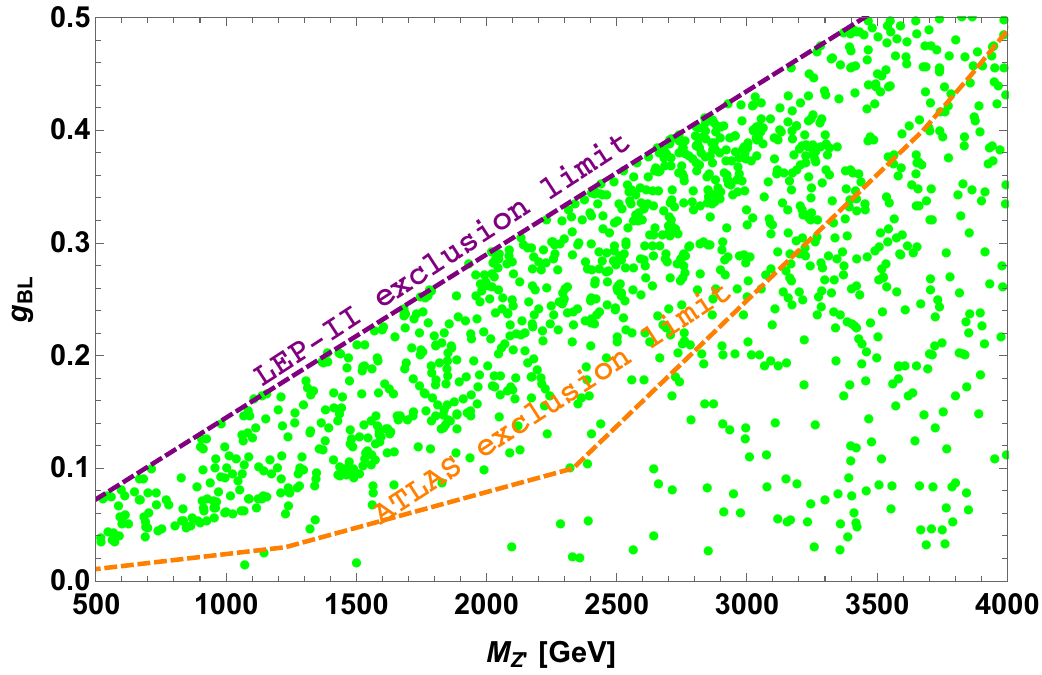}
\caption{ATLAS dilepton constraints on the proposed model are shown.  In the left panel, the black
	dashed line represents the exclusion limit from ATLAS \cite{ATLAS-CONF-2015-070}, while the colored lines represent the
	dilepton signal cross sections for different values of $g_{BL}$ as a function of $M_{Z^{\prime}}$.   The right panel shows ATLAS and LEP-II
	exclusion limits from dilepton searches in the plane of $M_{Z^{\prime}} - g_{BL}$.}
\label{ATLAS}
\end{center}
\end{figure}

For the present $B-L$ model, we use CalcHEP \cite{Belyaev:2012qa,Kong:2012vg} to compute the
production cross section of $Z^{\prime}$
\footnote{The more on LHC sensitivities in this class of $B-L$ model was recently performed in
	refs~\cite{Klasen:2016qux,Okada:2016gsh} }. Working in the mass range of $M_{Z^{\prime}} \le 4 $ TeV, we show in  the left panel of Fig.
\ref{ATLAS}, dilepton ($ee,\mu\mu$) signal in $Z^{\prime}$ production as a function of
$M_{Z^\prime}$.  It can be seen that for $g_{BL}=0.4$, the region below $M_{Z^{\prime}}\simeq3.7$ TeV
is excluded while for $g_{BL}=0.1$, $M_{Z{^\prime}}<2.3$ TeV is excluded.  Thus, for
$g_{BL}\gtrsim0.1$ the parameter space is pushed to heavier $M_{Z^{\prime}}$ above $2.3$ TeV.  For $g_{BL}<0.03$ we have
$M_{Z^{\prime}}\gtrsim 1.2$ TeV and for $g_{BL}=0.01$ we have $M_{Z^{\prime}}\gtrsim 0.5$ TeV.  We
see that the dilepton signal in $Z^{\prime}$ decay can impose stringent constraints on these models.  The right panel in Fig. \ref{ATLAS}
describes the parameter space in $M_{Z^{\prime}}-g_{BL}$ plane consistent with the current $3\sigma$
limit on relic density from PLANCK \cite{Ade:2015xua}. The region to the right of both the curves is consistent with ATLAS \cite{ATLAS-CONF-2015-070} and LEP-II  \citep{Schael:2013ita} bounds.
With ATLAS limit being the most stringent one, from the plot one can see that the model still has
a significant portion of the
parameter space that can satisfy the relic density. Thus, in general, we conclude that dilepton searches from LHC in
$Z^{\prime}$ models can pose stringent limits on the parameter space.  
\section{Light neutrino mass}

%================================================
\begin{table}[tb!]
\begin{center}
%================================================
\begin{tabular}{|c|c|c|c|}
	\hline
		~ Field~	& ~$SU(2)_L\times U(1)_Y$	~& ~$U(1)_{B-L}$~\\
	\hline

			 $\eta$		& $(\textbf{2}, 1/2)$	&   $-3$\\  
%			 & $\eta $ & 	 $(\textbf{2}, 1/2)$ & $-3 $\\
%			& $\Delta $  		& $(\textbf{3}, 1)$ & $-6 $\\
	\hline
	\hline
\end{tabular}
\caption{Inert doublet and its charge assignment.}
\label{tab:New_BL_DM}
\end{center}
\end{table}
Since the current model doesn't contain the right-handed neutrinos, the standard type-I
seesaw mechanism to generate light neutrino mass is not feasible with the existing particle content.  However, the neutrino masses
  can be generated at one-loop level through radiative mechanism, which will be briefly described in this section. For this purpose, we Introduce  an additional  inert doublet $\eta = \begin{pmatrix}
		 \eta^+		\\
		 \frac{S+iA}{\sqrt{2}}	\\
	\end{pmatrix}$ with the $B-L$ charge $-3$. Thus, the trivial scalar potential gets modified with the inclusion of  additional terms given as 
 \begin{eqnarray}
V^{\prime}&=& V(H,\phi_1,\phi_8) + \mu_{\eta}(\eta^{\dagger}\eta)+ \lambda_{\eta}(\eta^{\dagger}\eta)^2 + \lambda'_{H\eta}(H^{\dagger}\eta)(\eta^{\dagger}H) + \frac{\lambda_{\eta 18}}{2\Lambda^3}\left[(H^{\dagger}\eta)^2 \phi_8\phi_1^2 + {\rm h.c.}\right] \nn\\
&+& (\eta^{\dagger}\eta)\left[\lambda_{H\eta}(H^{\dagger}H) + \lambda_{\eta1}(\phi_1^{\dagger}\phi_1) +\lambda_{\eta8}(\phi_8^{\dagger}\phi_8)\right],
 \end{eqnarray}
where $\Lambda$ is the cut-off parameter. The masses of real and imaginary components of the inert doublet $\eta$ are  given as
\begin{eqnarray}
M_{S}^2 &=& \mu_{\eta}^2  +  \frac{ \lambda_{\rm \eta1}}{2} v^2_1 + \frac{ \lambda_{\eta8}}{2} v^2_8 +  \left(\lambda_{H\eta} + \lambda'_{H\eta}\right)\frac{v^2}{2}  + \lambda_{\eta 18}\frac{v^2 v_1^2v_8}{4\sqrt{2}\Lambda^3}, \nn\\
M_{A}^2 &=& \mu_{\eta}^2  +  \frac{ \lambda_{\rm \eta1}}{2} v^2_1 + \frac{ \lambda_{\eta8}}{2} v^2_8 +  \left(\lambda_{H\eta} + \lambda'_{H\eta}\right)\frac{v^2}{2} - \lambda_{\eta 18}\frac{v^2 v_1^2v_8}{4\sqrt{2}\Lambda^3}.
\end{eqnarray}
\begin{figure}[htb]
\begin{center}
\includegraphics[width=7 cm,height= 5.0 cm, clip]{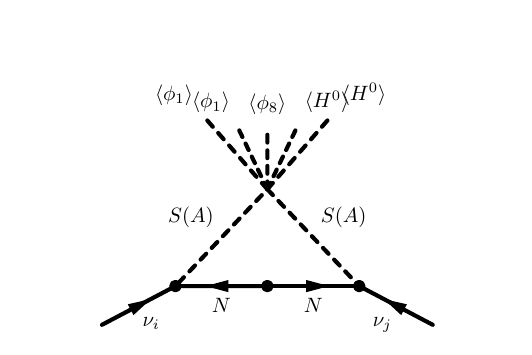}
\caption{Radiative generation of neutrino mass}
\label{rad_nu}
\end{center}
\end{figure}
With these particle content, one can write the interaction term to generate light neutrino mass at one-loop level as shown in Fig. \ref{rad_nu} as
\begin{equation}
\sum_{\alpha=1,2} Y_{i\alpha} \overline{(\ell_L)}_i \tilde{\eta} N_{\alpha R}.
\end{equation}
Thus,  from Fig. \ref{rad_nu}, one can write the light neutrino mass matrix \cite{Ma:2006km}, as
\begin{equation}
({\cal M}_\nu)_{ij} =  
\sum_{\alpha=1}^2 {Y_{i \alpha} Y_{j\alpha} M_{D\alpha} \over 16 \pi^2} \left[ 
\frac{M_S^2}{M_S^2- M_{D\alpha}^{2}} \ln \frac{M_S^2}{M_{D\alpha}^{2}} - 
\frac{M_A^2}{M_A^2- M_{D\alpha}^{2}} \ln \frac{M_A^2}{M_{D\alpha}^{2}} \right].
\end{equation}
Here
$M_{D\alpha} = (U^T M_R U)_{\alpha}$ and $N_{D\alpha} = U^{\dagger}_{\alpha \beta} N_{\beta}$, with $M_R$ being the Majorana mass matrix. 
If we assume $m_0^2 = (M^2_S + M^2_A)/{2}$ is much greater than $M^2_S -M^2_A = \frac{\lambda_{\eta 18}}{2\sqrt{2}\Lambda^3}v^2 v_1^2v_8$, the expression for the radiatively generated neutrino mass becomes
\begin{equation}
({\cal M}_\nu)_{ij} = {\lambda_{\eta 18} v^2v_1^2v_8  \over 32 \sqrt{2}\pi^{2}  \Lambda^3} 
\sum_{\alpha=1}^2 {Y_{i \alpha} Y_{j\alpha} M_{D\alpha} \over m_0^{2} - M_{D\alpha}^{2}} \left[ 
1 - {M_{D\alpha}^{2} \over m_0^{2}-M_{D\alpha}^{2}} \ln {m_0^{2} \over M_{D\alpha}^{2}}  \right].
\end{equation}
We further assume that inert doublet components are heavier than the DM mass.  Note that for the parameter space considered here, the range of cutoff scale $\Lambda$, which is allowed by perturbative limits is $\sim [50,10^4]$ TeV. For example, with $(Y,\lambda_{\eta 18}) \sim (10^{-1},10^{-2})$ and $(v_1,v_8,m_0,M_{D\alpha},\Lambda) \sim (2,2,2,0.5,100)$ TeV, one can have $m_{\nu} \sim 10^{-11}$ GeV.  
Thus, the light neutrino  mass generation can be successfully achieved in the proposed model.
% Since the above mechanism for
% neutrino mass generation does not affect the DM phenomenology, which is focus of the
% present work we shall not be presenting a detailed analysis of neutrino masses in this
% model.
\section{Conclusion} In this article, we made a detailed study of Majorana dark matter in
a variant of $B-L$ model where the gauge symmetry is extended with a $U(1)_{B-L}$. The
current model is enriched with three exotic fermions with $B-L$ charges $-4, -4, +5$, to
avoid the triangle gauge anomalies. The scalar sector is equipped with two additional
scalar singlets $\phi_1$ and $\phi_8$ with $B-L$ charges $-1,+8$ to break the $U(1)_{B-L}$
gauge symmetry giving mass to the exotic fermions and the heavy gauge boson $Z^{\prime}$.
The structure of the model is extremely fruitful, giving two kinds of mediators that
connect the visible and dark sectors. The lightest mass eigenstate upon the
diagonalization of exotic fermion mass matrix, plays the role of dark matter.  The scalar portal relic
abundance has been studied with all possible annihilation channels and the effect of massless physical Goldstone boson  is suitably addressed. The SI
cross section has been calculated and investigated with the current limits from LUX
(2016), XENON1T (2017) and PandaX (2017). Similar strategy is repeated for
$Z^{\prime}$-portal channels. But in the $Z^{\prime}$ case, it is not possible to study
for direct searches as the Majorana dark matter couples axial-vectorially with the
$Z^{\prime}$, while SM quarks couple to $Z^{\prime}$ vectorially.  In collider searches,
the ATLAS bounds on the $Z^{\prime}$ mass and $g_{BL}$ impose strong constraints.
However, we still have a viable parameter space satisfying the current relic density and
the dilepton bounds. 
We have also addressed the generation of light neutrino mass by adding an additional inert doublet $\eta$ with $B-L$ charge assigned as $-3$. To conclude, we have made a complete systematic study
of Majorana dark matter in a new variant of $B-L$ gauge extended model. This simple model
survives the current collider limits while satisfies dark matter constraints and can be
probed in future high luminosity data from LHC.
%an example for the experimentalists to decide on the frameworks to be emphasized in the
%quest for dark matter.

\acknowledgments 
SS would like to thank Dr. Subhadip Mitra for the help in CalcHEP code and Department of Science and Technology (DST) - Inspire Fellowship division, Govt of India for the financial support through ID No. IF130927.  RM would like to thank Science and Engineering Research Board (SERB), Government
of India for financial support through grant No. SB/S2/HEP-017/2013.

\bibliographystyle{utcaps_mod}
\bibliography{BL}

\end{document}